\documentclass[prl,aps,amsmath,amssymb,twocolumn,superscriptaddress]{revtex4-2} 

\usepackage[colorlinks=true,citecolor=blue]{hyperref}
\usepackage{graphicx}
\usepackage{amsmath, amstext, amssymb, amsfonts, amsxtra}
\usepackage{xspace}
\DeclareSymbolFontAlphabet{\amsmathbb}{AMSb} 
\usepackage{xcolor}
\usepackage{braket}
\usepackage[normalem]{ulem}

\newcommand{\HL}{H_{\rm L}}
\newcommand{\HR}{H_{\rm R}}

\newcommand{\NL}{N_{\rm L}}
\newcommand{\NR}{N_{\rm R}}

\newcommand{\epsL}{\varepsilon^{\rm L}}

\newcommand{\cur}{\mathcal{I}}

\newcommand{\Imicper}{\cur^{\rm mic}_{\rm perturb}}

\newcommand{\tr}{\mathrm{Tr}}

\newcommand{\im}{\mathrm{i}}

\newcommand{\BL}{B^{\rm L}}
\newcommand{\dBL}{\dot{B}^{\rm L}}

\newcommand{\BR}{B^{\rm R}}

\newcommand{\sx}[1]{\sigma^{x}_{#1}}
\newcommand{\sy}[1]{\sigma^{y}_{#1}}
\newcommand{\sz}[1]{\sigma^{z}_{#1}}

\newcommand{\para}[1]{{\em #1}\/.---}

\newcommand{\sutd}{Science, Mathematics and Technology Cluster, Singapore University of Technology and Design, 8 Somapah Road, 487372 Singapore} 
\newcommand{\sutdepd}{EPD Pillar, Singapore University of Technology and Design, 8 Somapah Road, 487372 Singapore} 
\newcommand{\hnu}{Key Laboratory of Low-Dimensional Quantum Structures and Quantum Control of Ministry of Education, Department of Physics and Synergetic Innovation Center for Quantum Effects and Applications, Hunan Normal University, Changsha 410081, China
}
\newcommand{\ictp}{The Abdus Salam International Centre for Theoretical Physics, Strada Costiera 11, 34151 Trieste, Italy}

\begin{document}

\title{Emergence of steady currents due to strong prethermalization} 

\author{Xiansong Xu} 
\affiliation{\sutd}
\author{Chu Guo} 
\affiliation{Henan Key Laboratory of Quantum Information and Cryptography, Zhengzhou,
Henan 450000, China}
\affiliation{\hnu}
\author{Dario Poletti}
\affiliation{\sutd} 
\affiliation{\sutdepd} 
\affiliation{\ictp}

\begin{abstract} 
A steady current between baths is a manifestation of the prethermalization phenomenon, a quasi-equilibrium dynamical process with weak conserved quantity breaking. 
We consider two finite nonintegrable many-body baths each following the eigenstate thermalization hypothesis, and each prepared in a random product state with fixed and different energy constraints, i.e., within the mean energy ensemble.
Such an initialization, not being constrained to superpositions or mixtures of many-body eigenstates, opens the door to experimental realization and also significantly simplifies numerical simulations.
We show that such dynamical process is typical as the current variance decreases exponentially with respect to the size of baths. We also demonstrate that the emerging current is prethermalized in a strong sense, analogously to strong thermalization, meaning that the current values stay close to the microcanonical one for most of the time.
\end{abstract}

\maketitle

\para{Introduction} The understanding of equilibration and thermalization of isolated quantum systems has been significantly advanced thanks to the renewed interest in the foundations of statistical mechanics. In particular, the emerging ``pure state quantum statistical mechanics'' has gradually become a mainstream paradigm to address the apparent disconnection between microscopic reversibility and macroscopic irreversibility \cite{BorgonoviZelevinsky2016, DAlessioRigol2016, GogolinEisert2016, MoriUeda2018, Deutsch2018, Ueda2020}. An important question in many-body systems is how accurately one needs to prepare a system in order to observe emerging equilibrium or nonequilibrium phenomena such as thermalization or emergence of steady currents. Regarding the occurrence of thermalization, pure state quantum statistical mechanics descriptions include concentration of measure \cite{LedouxLedoux2001, MullerEisert2011}, typicality \cite{GoldsteinZanghi2006,PopescuWinter2006, Reimann2007, BartschGemmer2009, Reimann2018a} and eigenstate thermalization \cite{Deutsch1991, Srednicki1994}. Importantly, thermalization is deemed to occur in a {\it weak} sense when the local observables converge towards their corresponding thermal equilibrium values after taking long-rime averages, while in a {\it strong} sense when convergence does not requires taking time averages \cite{BanulsHastings2011,DAlessioRigol2016,LinMotrunich2017,ChenPan2021}. 

Meanwhile, nonequilibrium dynamics also features stationary behaviors such as the steady current flow, and their emergence can be studied also from a pure-state statistical mechanics perspective. Considering two finite-size ``baths'' coupled to each other, each initially prepared in mixtures or superpositions of eigenstates, it has been possible to show that steady currents can emerge for long times \cite{MonnaiYuasa2014, MonnaiYuasa2016}, and in particular that they result from a prethermalization mechanism \cite{XuPoletti2022}. Other attempts include the generalization of ETH to nonequilibrium systems through effective Hamiltonians or temperatures \cite{MoudgalyaSondhi2019a, ShiraiMori2020}, and similar bipartite systems for transport and thermalization were studied in Refs. \cite{BushongDiVentra2005, PonomarevHanggi2011, BiellaFazio2016, MascarenhasSavona2017, LjubotinaProsen2017, ZnidaricLjubotina2018, BalachandranPoletti2018b, LjubotinaProsen2019,BiellaMazza2019, PulikkottilTomsovic2020, PulikkottilTomsovic2022}.     

We ask ourselves if the same concepts of strong and weak emergence occur also for nonequilibrium processes, with particular emphasis on the emergence of steady current formations. In addition, we would like to decouple such an investigation from the ability to compute, or experimentally prepare, eigenstates of the Hamiltonian of the bath, but only requiring that the expectation value of the energy of each bath is in a particular energy window. This significantly expands previous results and allows experimental realizations, for instance with ultracold atoms, ion traps, or solid states setups \cite{BlochZwerger2008, GeorgescuNori2014, MonroeYao2021, BluvsteinLukin2021, EbadiLukin2021}.

In this work we consider two non-integrable spin chains as finite-size baths, and we prepare them in a tensor product of single-site random states.  We first show that these random product states, with respect to each spin chain, are typical, i.e., a single sample well represents an ensemble of such states subject to an energy constraint (within the mean energy ensemble) \cite{BenderHook2005, BrodyHughston1998, JiFine2011, MullerEisert2011,FreschMoro2013, BrodyHughston2016,  GogolinEisert2016}. As a consequence, we then demonstrate that, once the spin chains are weakly coupled at one edge, a typical energy current is formed between the two chains. The energy current can be regarded as typical as it converges towards the prediction from microcanonically prepared baths, and the variance of currents decreases exponentially with the size of the baths. We achieve this by studying systems up to 52 spins with an accurate perturbative method which agrees with the results accessible by exact simulation in the weak coupling limit. 
More importantly, we show that the emergent prethermalized steady current has, at each point in time, a variance that decreases exponentially with the system size without the need to take any time averages. Hence we are able to demonstrate that the weak coupling currents flow is indeed steady due to {\it strong prethermalization}. 

\para{Model}
Two finite-size quantum systems $\HL$ and $\HR$ are treated as the left and right baths, which are coupled via an interaction term $V$, i.e., the total Hamiltonian of the system is $H = \HL + \HR + V$. 
Each bath is taken to be a nonintegrable spin chain. The Hamiltonian of the left bath is given by 
\begin{align}
  \HL = \sum_{n=1}^{N_{\rm L}-1} J_{zz} \sz{n}\sz{n+1} + J_{yz} \sy{n}\sz{n+1} + \sum_{n=1}^{N_{\rm L}}  h_x \sx{n} + h_z \sz{n}, 
  \label{eq: hamiltonian}
\end{align}
while the right bath has the same Hamiltonian with the site index $n$ starting from $N_{\rm L} + 1$ to $N=N_{\rm L} + N_{\rm R}$. It follows that $\NL$ and $\NR$ denote the size of the two baths. The interaction between them is given by $V = \gamma \BL \bigotimes \BR$ where $\gamma$ is the coupling strength and $\BL$ and $\BR$ are the operators acting on the left and right baths, respectively. In the following, we consider $\BL=\sigma^x_{\NL}$ and $\BR=\sigma^x_{\NL+1}$ and we use $J_{yz}=J_{zz}$, $h_x=-1.05J_{zz}$, $h_z=0.5J_{zz}$ such that each bath is chaotic with Gaussian unitary ensemble level statistics \cite{AtasRoux2013}. We note that such a choice of the model has been used in Refs. \cite{XuPoletti2022, HuangZhang2019}. In the following, we work in units such that $J_{zz}=\hbar=1$.   

\para{Energy currents}
The only conversed quantity of each bath, when isolated, is the total energy. We can thus define the energy current operator with respect to the left bath as 
\begin{align}
  I^{\rm L}  = \frac{d\HL}{dt}=-\im [V, \HL] = \gamma  \dBL \otimes \BR,
  \label{eq: current_operator}
\end{align}
where we have defined $\dBL \equiv -\im \left[\BL,\HL\right]$. To see a clear relation between variance of current and bath size, we push the simulation to 52 spins with a perturbative expression for the current given by \cite{ZhouZhang2020, XuPoletti2022} similar to the master equation formalism \cite{EspositoGaspard2003a, EspositoGaspard2007, ThingnaWang2014, Riera-CampenyStrasberg2021, LandiSchaller2021}. 
\begin{align}
  \cur(t)  = & \gamma \mathcal{\dot{B}}^{\rm L}(t) \mathcal{B}^{\rm R}(t) - \im \gamma^2 \int_{0}^{t} d \tau
  {\Big [ } \overrightarrow{\mathcal{C}}_{\rm L}(t,\tau) \mathcal{C}_{\rm R}(t,\tau) \nonumber\\ 
             &-\overleftarrow{\mathcal{C}}_{\rm L}(\tau,t) \mathcal{C}_{\rm R}(\tau,t) {\Big ]}, 
             \label{eq: current}
\end{align}
where $\mathcal{\dot{B}}^{\rm L}(t)=\tr_{\rm L}\left[\rho_{\rm L} \dBL(t) \right]$, $\mathcal{B}^{\rm R}(t)=\tr_{\rm R}\left[\rho_{\rm R} \BR(t) \right]$ and we have defined the two-time correlation functions
\begin{align}
  \overrightarrow{\mathcal{C}}_{\rm L}(t,\tau) &= \tr_{\rm L}\left[\rho_{\rm L} \dBL(t)\BL(\tau)\right], \label{eq: twotimecorrL1}\\
  \overleftarrow{\mathcal{C}}_{\rm L}(t,\tau) &= \tr_{\rm L}\left[\rho_{\rm L} \BL(\tau)\dBL(t)\right], \label{eq: twotimecorrL2}\\ 
  \mathcal{C}_{\rm R}(t,\tau) &= \tr_{\rm R}\left[\rho_{\rm R} \BR(t)\BR\right].
  \label{eq: twotimecorrR}
\end{align}
Here we have also defined $B^{\rm L/R}(t)=e^{\im H_{\rm L/R} t}B^{\rm L/R}e^{-\im H_{\rm L/R} t}$ and the same applies to $\dBL(t)$. For more details on the numerical computations see \cite{Supp}.

\begin{figure}
  \includegraphics[width=\columnwidth]{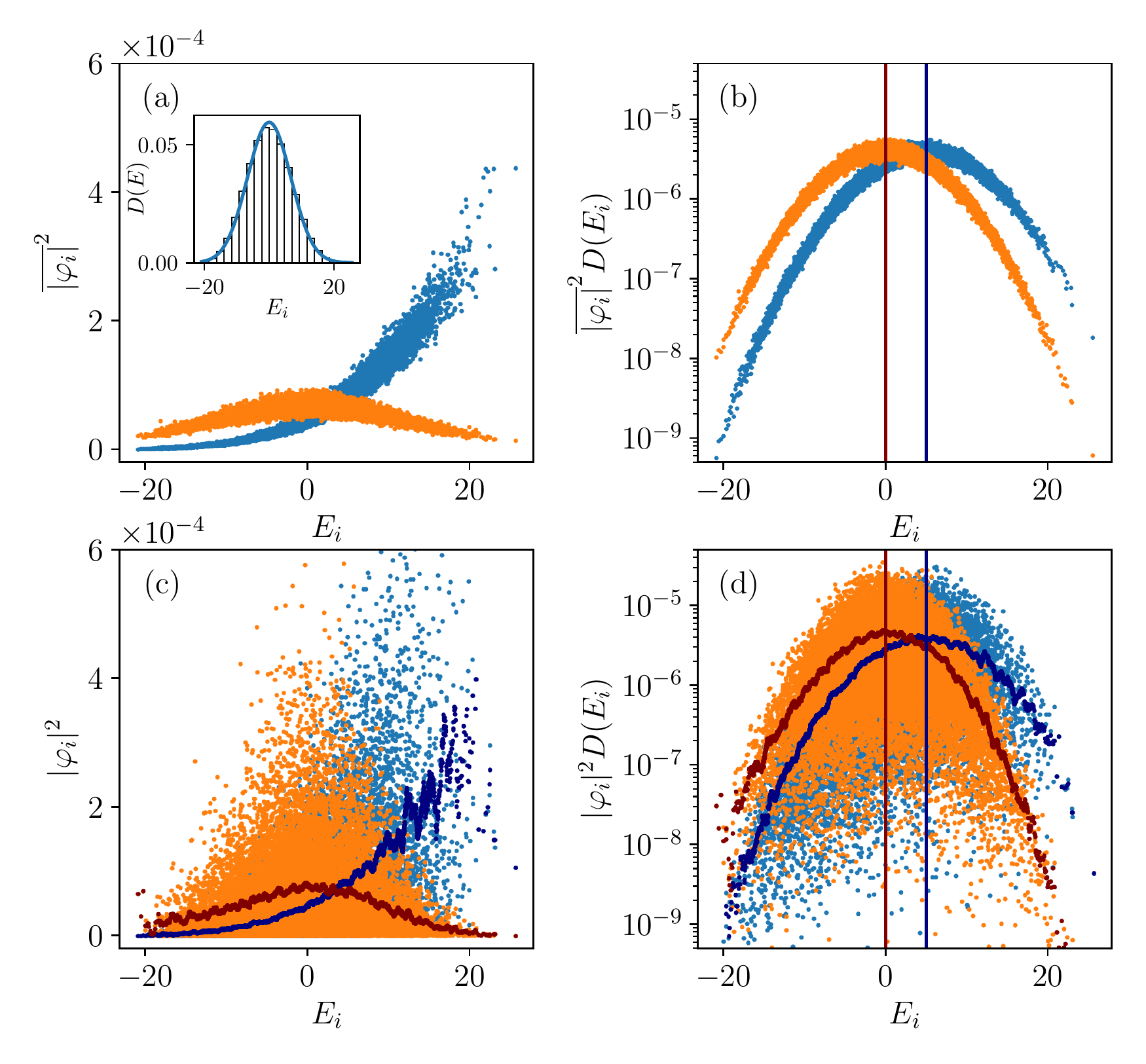}
  \caption{Initial states profile for $N_{\rm L}=14$ for (a, b) 100 samples and (c, d) single sample. The left panels denote the (averaged) probability distribution in the eigenbasis and the right panels show the shifted distribution with a fitted many-body density of states shown in the inset of (a). In (c,d), the brown and darker blue curves denote the moving average of the single initial state with a window $\delta E = 0.5$. The initial states are chosen to be random product states with the energy constraint given by $\epsL=5/12$. 
  }
  \label{fig: state_analysis}
\end{figure}

\para{Preparation and properties of the baths} We consider a class of generic initial states $\{\ket{\psi_{\alpha}}\}$ with the constraint that their energies $E^{\alpha} = \braket{ \psi_{\alpha} | H_{\alpha} | \psi_{\alpha}} $ is subject to the target energies $\mathcal{E}_{\alpha} = \varepsilon_{\alpha} N_{\alpha}$ where $\varepsilon_{\alpha}$ is the energy per site. Here $\alpha={\rm L,R}$ refer to the left or right bath. More precisely, the initial states of a bath are chosen such that
\begin{align}
  | \braket{ \psi_{\alpha} | H_{\alpha} | \psi_{\alpha}}  - \mathcal{E}_{\alpha} | & < \eta_{\alpha},
  \label{eq: econstraint}
\end{align}
where $\eta_{\alpha}$ are small real numbers. Throughout the work, we consider $\eta_{\rm L} = \eta_{\rm R} = 0.01$. 

For the ease of experimental considerations, we consider the initial states of each bath as single random product states
\begin{align}
  \ket{\psi_{\alpha}} = \bigotimes_n \left( u_{\alpha,n} \ket{\uparrow}_n + v_{\alpha,n} \ket{\downarrow}_n    \right),  \label{eq: ansatz}
\end{align}
where $|u_{\alpha,n}|^2 + |v_{\alpha,n}|^2=1$ and $u_{\alpha,n}$ and $v_{\alpha,n}$ are sampled, or optimized, such that they satisfy the energy constraint defined in Eq. (\ref{eq: econstraint}).

To obtain the initial states of the baths efficiently, we substitute Eq.~(\ref{eq: ansatz}) as a variational ansatz (randomly initialized parameters) into Eq.~(\ref{eq: econstraint}), and use a gradient-based (L-BFGS) minimization algorithm \cite{GuoPoletti2021}. We stress here that such initial states are, in general, superposition of many eigenstates with very different energies. These initial states can also be easily implemented on quantum computers using a single layer of rotational gates, and more generally in ion traps, superconducting qubits setups ,and ultracold atoms \cite{BlochZwerger2008, GeorgescuNori2014, MonroeYao2021, BluvsteinLukin2021, EbadiLukin2021}. 

We now show that, for each isolated bath, such a random product initial state, Eq.~(\ref{eq: ansatz}), is ``typical'' in the sense that a single random sample reproduces the features of the average. We first consider the property of the averaged wavefunction. In Fig. \ref{fig: state_analysis}, the top panels (a, b) demonstrate the probability $|\varphi_i|^2$ of the prepared state in the eigenbasis, averaged over 100 sampled states prepared with energy per site $\varepsilon_{\rm L}=5/12$ (blue) and $0$ (orange). While these distributions may not seem particularly insightful, their role becomes clear when weighted with the density of states $D(E)$ where $E$ is the energy, as shown in Fig. \ref{fig: state_analysis}(b). This shows that these random product states have a weighted distribution peaked around the desired energy constraints, as indicated by the navy and maroon vertical lines. 
Furthermore, if we now consider a single initial condition represented by orange and blue dots in Fig. \ref{fig: state_analysis}(c,d), not averaged over 100 as in the panels (a,b) and we consider moving averages \footnote{a moving average defined as 
  $\Phi(E) = \sum_{i;|E_i - E| \le \delta E} |\varphi(E_i)|/M $
where $M$ is the number of eigenstate that satisfies $|E_i -E| \le \delta E|$. With an energy width of $\delta E=0.25$} of $|\varphi_i|^2$ and $D(E_i)|\varphi_i|^2$ denoted by the maroon and navy dots, we see a clear correspondence appears between the curves in the top and bottom panels. This shows that each initial state can represent the properties of the ensemble.

Another important step we perform to improve the quality of the results is that once initialized the baths, we let them evolve for a preparation time $\tau_{\rm prep}$ such that the local observables relax close to a single value, i.e., thermalize. This $\tau_{\rm prep}$ has been chosen such that it is larger than the intrinsic time which the bath takes to thermalize, $\tau_{\rm thermal}$. In the computations done we use $\tau_{\rm prep} = 50$, which is comfortably sufficient to observe thermalization within the baths before they are coupled, and it also allows the emergence of quasi-steady currents shortly after we couple them. For more details about this point, see \cite{Supp}.

\begin{figure*}
  \includegraphics[width=\textwidth]{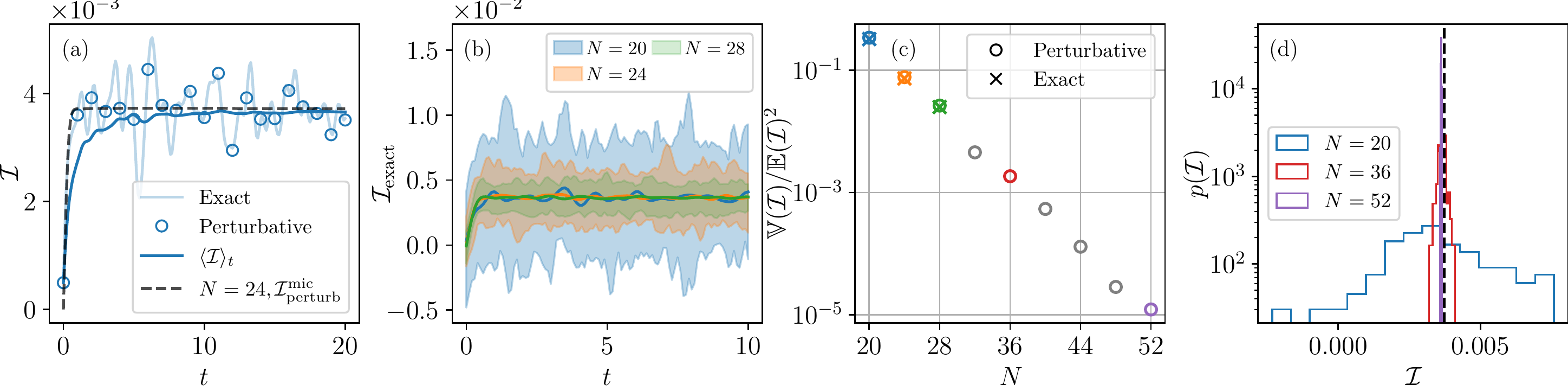} 
  \caption{(a) Current dynamics for exact (light blue line), perturbative (blue markers), and time-averaged (solid blue line) for $N=28$. The dashed line denotes the perturbative microcanonical current $\Imicper$ for $N=24$ with an energy shell width $0.1$. (b) Fluctuations of the exact energy current for different total system size $N=20$ (blue shades), $24$ (orange shades), $28$ (green shades) with 100 samples. The shaded regions are determined by the extrema of the energy current at each time. The coupling strength is $\gamma=0.1$ and the energy per spin for the baths is $\varepsilon_{\rm L} = -\varepsilon_{\rm R}=5/12$. 
  (c) Variance of the energy currents with different system sizes at $t=5$. The variances for exact current are given by the cross markers for up to $N=28$ and the variances for perturbative results are given by empty circle markers.
(d) Histogram of the perturbative currents at $t=5$ for $N=20$ (blue), $N=36$ (red), $N=52$ (purple). The vertical dashed line represents the perturbative microcanonical current value at $t=5$ for $N=24$ taken from (a).}
  \label{fig: typical_variance-exp}
\end{figure*}

\para{Typical quasi-steady currents} Having defined the initial states of the baths followed by a thermalization dynamics with time $\tau_{\rm prep}$, we are ready to explore the steady currents between them. We first consider a proof of principle calculation with a single initial condition for both exact and perturbative currents. In Fig. \ref{fig: typical_variance-exp}(a) each of the baths consists of 14 spins and thus the composite system contains $N=28$ spins. The baths are prepared with energy constraints given by $\varepsilon_{\rm L} = - \varepsilon_{\rm R}=5/12$. The exact energy current dynamics is given by the oscillating light blue line indicated in Fig. \ref{fig: typical_variance-exp}(a) complemented, with circles, by perturbative results using Eq. (\ref{eq: current}). The perturbative results are consistent with the exact results with their differences negligible at the order of magnitude of the current. It is thus possible to study the variance of the currents for larger systems using perturbative calculations.  
For the baths size $N_{\rm L/R}=14$, the exact current fluctuates and it might be difficult to be regarded as a steady current at the first glance. However, if we take the time-averaged currents starting from $t=0$, denoted as $\braket{\mathcal{I}}_t = (1/t)\int_0^t \mathcal{I}(\tau) d\tau$, we immediately see that the time-averaged current converges to an almost steady value, consistent with the perturbative microcanonical current denoted by the black dashed line computed for $N=24$. 
We remark that by microcanonical current we mean the value of the current expected from preparing each bath in a uniform mixture of eigenstates in the respective energy windows. The result depicted in Fig.~\ref{fig: typical_variance-exp} can be thought to be related to weak thermalization, for which $\braket{\mathcal{I}}_t \approx \braket{\mathcal{I}}_{\rm mic}$. We shall show that for larger bath sizes the system actually follows strong prethermalization. First, however, we are interested in the notion of typicality, i.e., whether at a given point in time the current from an initial state can represent the average current over many different initial states, all chosen with the same energy constraint.

To do this, we study how the variance of the currents changes with respect to the total system sizes. 
We first consider 100 different initial preparations of the baths. The shaded regions in Fig.~\ref{fig: typical_variance-exp}(b) are bounded by the maximum and minimum values of the currents considering all the different initial preparations for different system sizes, $N=20$ (blue), $24$ (orange), and $28$ (green). 
We observe a significant narrowing of the fluctuation as indicated by the reduced width of the shaded areas with the system size. Their time-averaged current values are given by the solid lines. To identify the scaling of the fluctuations with the system size, we apply the perturbative approach of Eq.~(\ref{eq: current}) for up to $52$ spins. In Fig. \ref{fig: typical_variance-exp}(c), by considering a fixed time $t=5$, we observe a clear exponential decay of the normalized variance $\mathbb{V}(\mathcal{I})/\mathbb{E}(\mathcal{I})^2$. The exact current, already shown to be consistent with perturbative results in Fig. \ref{fig: typical_variance-exp}(a), has consistent variance with the perturbative one, as clearly indicated by the overlapped markers. Last, in Fig. \ref{fig: typical_variance-exp}(d) we show the probability distribution of the currents over the 100 samples. This shows in a dramatic way that the variance of the current becomes very narrow. 
Furthermore, the peak of the distribution, considering the finite-size of systems studied, corresponds with the vertical dashed line in Fig. \ref{fig: typical_variance-exp}(d) which is the value of the current expected from a microcanonical preparation of the baths \cite{XuPoletti2022}.   
We can thus expect, for large enough systems, a vanishing variance of currents for different realizations of initial states. Also, a single random product state, with the sole constraint on energy, can be a representative state for any other (similarly built) initial state.

\begin{figure}
  \includegraphics[width=\columnwidth]{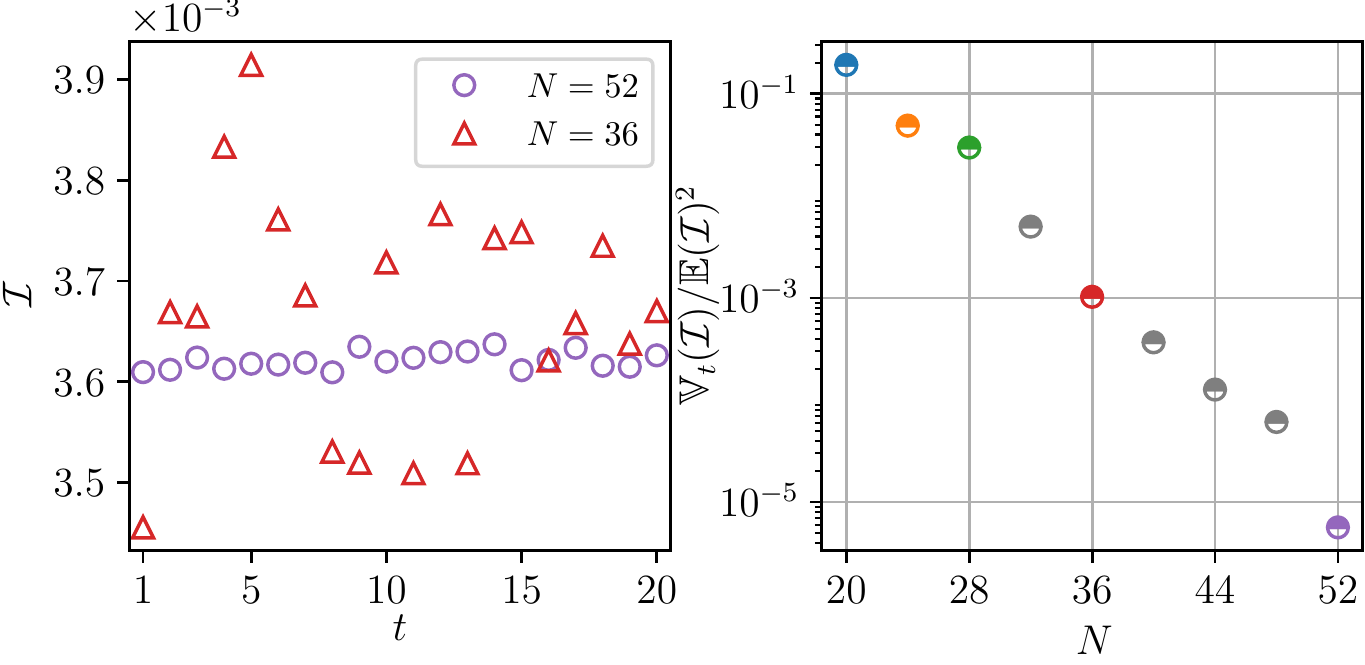}
  \caption{(a) Current dynamics versus time via perturbative methods for $N=10$, $N=36$, and $N=52$. (b) The variance of the energy current over time (from $t=1$ to $t=20$) versus different system sizes each for a single random product state. Other parameters are $\gamma=0.1$ and $\varepsilon_{\rm L}=-\varepsilon_{\rm R} = 5/12$.  }
  \label{fig: prethermalization}
\end{figure}

\para{Strong prethermalization dynamics of quasi-steady currents}
To show that the emerging currents are indeed steady, for each initial preparation, we study the dynamics of the current as a function of time. In Fig. \ref{fig: prethermalization}(a) we observe the (perturbative) currents for $N=36$ and $N=52$ versus time for a single realization of the baths. This shows that the fluctuations over time are significantly smaller for larger baths. To be more quantitative, in Fig. \ref{fig: prethermalization}(b) we show that the variance over time (measured from times $t=1$ to $t=20$ and denoted as $\mathbb{V}_t(\mathcal{I})$) decreases exponentially with the bath size. Fig. \ref{fig: prethermalization} clearly indicates that the steady current follows a ``strong thermalization'' behavior, i.e., the current magnitude is consistently close to the prethermal value which is expected, as shown from the dashed horizontal line, from microcanonically prepared baths, and such proximity to this value increases exponentially with the size of the baths. For dependence of the current on energy differences between the baths and characterization of the thermalizing long-time dynamics, see \cite{Supp}.

\para{Conclusions} We have studied the current dynamics between two finite-size nonintegrable many-body quantum systems which serve as baths. We prepare each bath in an initial random product state with an energy constraint. The sole presence of the energy constraint defines a set of typical initial conditions resulting in a typical nonequilibrium current dynamics, in the sense that any random initial state chosen with energy within that window would result in the same prethermal nonequilibrium quasi-steady current. This is demonstrated by converging mean current values and an exponentially decreasing variance of the current from different initial states chosen with the same energy constraints. 

Besides typicality, we have also studied if the current is truly steady and if the steadiness has to rely on the time averaging similar to the weak thermalization. More precisely, we found that the variance over time for the current during the prethermal regime for a single realization of the initial condition decreases exponentially with the bath size. Such an exponential decay of the variance over time, combined with the observation that the current approaches the expected value for a microcanonical preparation of the baths, allow us to conclude that the nonequilibrium steady dynamics follows strong prethermalization.  

The chosen initial states have two very important advantages compared to preparing superposition or mixtures of eigenstates as done in Refs. \cite{MonnaiYuasa2014, XuPoletti2022}. From a numerical point of view, we get rid of the computation of the eigenstates of the Hamiltonian and we can thus afford to study larger systems so that we can observe a clear signature of exponential decay of the variance of the currents (over samples or over time) with the system size. Furthermore, it would also be possible to use matrix product state techniques to prepare and evolve such states, although the bond dimension could still grow significantly thus making the simulations still challenging. The second advantage is that these initial conditions can be readily prepared in the state of the art experiments provided one can estimate the expectation value of the energy in the setup. 

A number of open questions remain to be pursued. For instance, one could consider the presence of a system between the baths and analyze how this affects the emergence of a quasi-steady current. Another important aspect is to study in detail the dynamics inside the baths as the quasi-steady current emerges and then eventually vanishes.  

\para{Acknowledgements} The authors are grateful to Zala Lenar\v{c}i\v{c} and Lev Vidmar for fruitful discussions. 
X. X. is grateful for the hospitality of Sichuan Normal University. D. P. acknowledges support from the Ministry of Education of Singapore AcRF MOE Tier-II (Project No. T2MOE2002). C. G. acknowledges support from National Natural Science Foundation of China under Grants No. 11805279, No. 61833010, No. 12074117 and No. 12061131011. The computational work for this article were partially performed on the National Supercomputing Centre, Singapore \cite{NSCC}.

\end{document}


\title{Supplementary Materials: \\Emergence of steady currents due to strong prethermalization}

\author{Xiansong Xu} 
\affiliation{\sutd}
\author{Chu Guo} 
\affiliation{Henan Key Laboratory of Quantum Information and Cryptography, Zhengzhou,
Henan 450000, China}
\affiliation{\hnu}
\author{Dario Poletti}
\affiliation{\sutd} 
\affiliation{\sutdepd} 
\affiliation{\ictp}

\date{\today}
\maketitle
\tableofcontents

\section{Thermalization of the bath of random product states \label{sec: unitary}}

\begin{figure}
  \includegraphics[width=\columnwidth]{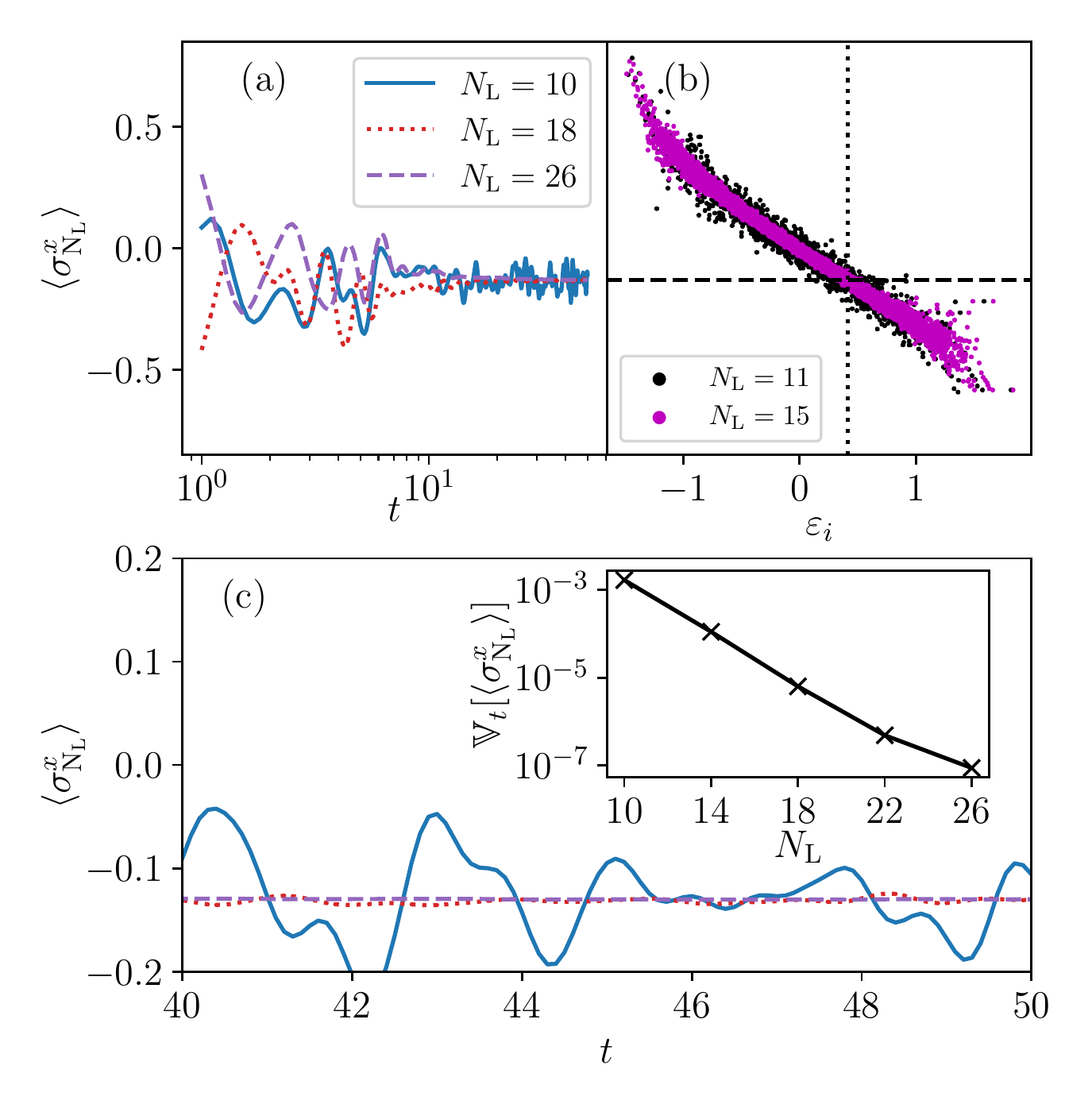}
  \caption{(a) Thermalization dynamics for $\langle \sigma^x_{N_{\rm L}}\rangle$ for different system size $N_{\rm L} = 10, 18, 26$ on a log scale up to $t=50$. (b) The local observable values for each eigenstates versus the corresponding energy per spin. The vertical dashed line denotes the initial energy per spin used in (a) and the horizontal dashed line denotes the microcanonical value of $\braket{\sigma^x_{N_{\rm L}}}$ for $N_{\rm L}=24$. (c) A zoomed in view of the thermalization dynamics between $t=40$ and $t=50$. The inset shows the variance over time of the $\langle \sigma^x_{N_{\rm L}}(t) \rangle$ from $t=40$ to $t=50$ for different system sizes up to $N_{\rm L} = 24$. The energy per spin is fixed to be $\varepsilon_{\rm L}=5/12$.}
  \label{fig: thermalization_dynamics}
\end{figure}

Since the baths are built such that they follow ETH and thus should, when left on their own, thermalize, we first show that the nonintegrable spin system we use can thermalize under unitary evolution when prepared as discussed in the main paper. Here, by thermalization, we refer to the strong thermalization notion where the observables converge toward the microcanonical averages for most of the time \cite{BanulsHastings2011, DAlessioRigol2016}.

In Fig. \ref{fig: thermalization_dynamics} we show that the evolution in time of the local observable $\braket{\sigma^x_{N_{\rm L}} (t)}$, i.e. the transverse spin of the last site, for the left bath when decoupled from the right bath. In Fig. \ref{fig: thermalization_dynamics}(a) we observe that this local observable converges in time towards a value with oscillations which become significantly smaller as we increase the size of the bath. In Fig. \ref{fig: thermalization_dynamics}(c) we zoom in the longer time dynamics and we can see very clearly how minuscule are the oscillations even for just $26$ spins. The inset shows the variance of the oscillations in a log-lin plot, showing an exponential decrease of the variance with the bath's size. 
In Fig. \ref{fig: thermalization_dynamics}(b) we show the diagonal elements of the operator $\sigma^x_{N_{\rm L}} $ in the energy basis. As expected from ETH, the values fall closer and closer to a continuous function of the energy per spin $\varepsilon_i$, which we refer to as $f(\varepsilon)$, as the bath size is increased from $N_{\rm L}=11$ to $N_{\rm L}=15$. The energy constraint is such that the energy per spin for different system sizes is fixed, i.e. $\varepsilon_{\rm L}=5/12$ and the total energy $E_{\rm L} = \NL \varepsilon_{\rm L}$. More importantly, in panel (b) we plot vertical dotted line at the energy given by the expectation value of the energy at which we have prepared the bath. We find that the asymptotic value of $\braket{\sigma^x_{N_{\rm L}} (t)}$ shown in Fig. \ref{fig: thermalization_dynamics}(a) corresponds to the horizontal dashed line in Fig. \ref{fig: thermalization_dynamics}(b) given by an microcanonical averaged value.

This is a strong indication of thermalization in the strong sense, where the observable values stay near the equilibrium values with negligible fluctuations. 

\section{Absence of preparation dynamics: competition of the timescale \label{sec: preparation}}

\begin{figure}
  \includegraphics[width=\columnwidth]{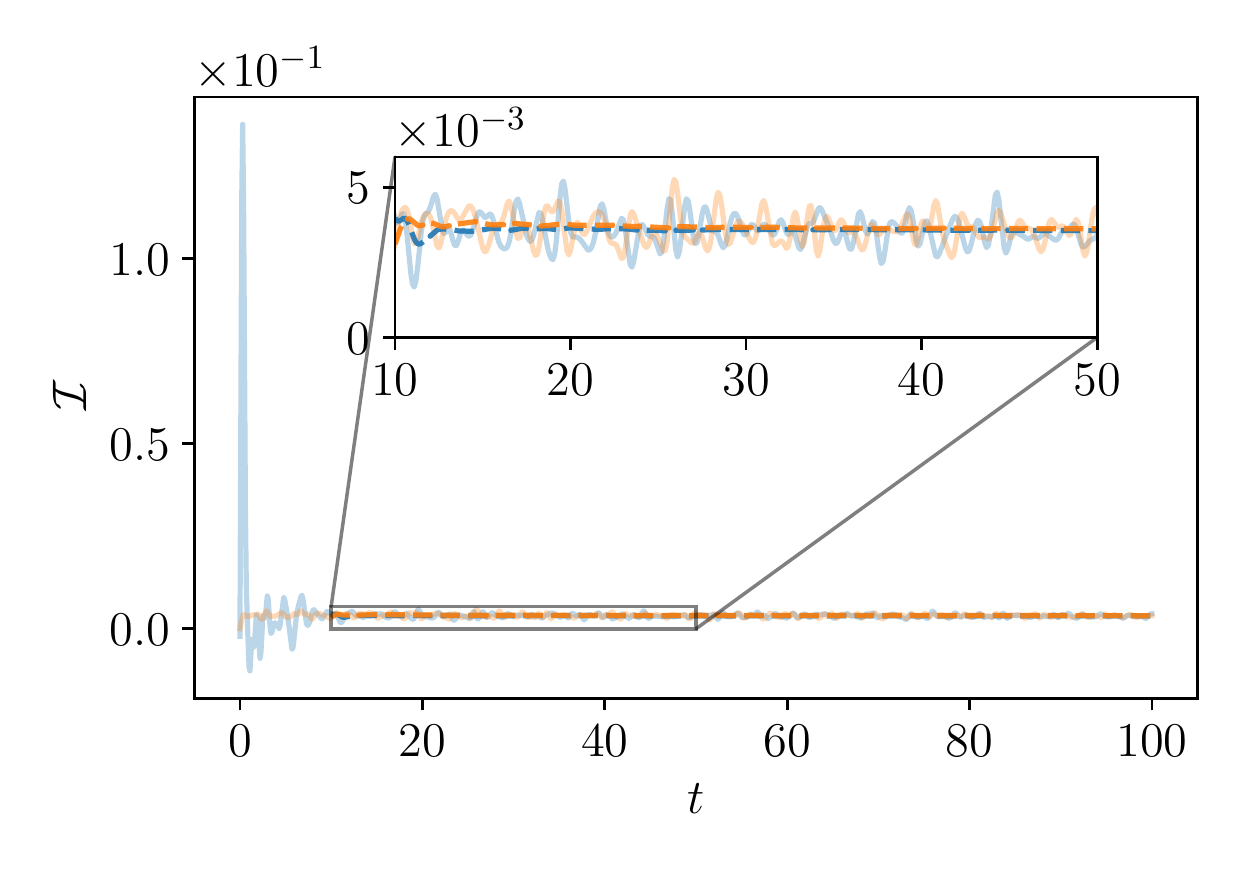}
  \caption{Comparison of exact current with a preparation dynamics of $\tau_{\rm prep}=50$ (orange solid line) and without preparation dynamics $\tau_{\rm prep}=0$ (blue solid line) for $N=28$. Their reverse time average are shown by the dashed lines respectively. Parameters are $\gamma=0.1$ and $\epsL=-\epsR=5/12$.}
  \label{fig: absence_preparation}
\end{figure}

In the main paper we have used a finite, and long enough, preparation time $\tau_{\rm prep}$ in which the baths are left to thermalize before being coupled to each other. If, however, the intrinsic thermalization time of the bath $\tau_{\rm thermal}$ is short, then one may not even need to allocate a time $\tau_{\rm prep}$ for the isolated evolution of the baths before coupling them. This is particularly relevant for our study in which the Hamiltonian of the baths allows a fast thermalization. In particular we have baths which thermalize over time scales shorter than the prethermalization time scale $\tau_{\rm prethermal}$, which is the time, in presence of weak coupling between the baths, required to observe the emergence of the quasi-steady currents. In short, we need that $\tau_{\rm prethermal} > \tau_{\rm thermal}$. 

In Fig.~\ref{fig: absence_preparation} we show the current obtained for $\tau_{\rm prep}=50$ (orange continuous line) and for $\tau_{\rm prep}=0$ (blue continuous line), and their respective cumulative averages (dashed lines in the respective colors). While the orange curve for $\tau_{\rm prep}=50$ reaches steady values much earlier, we observe a very clear convergence of both the cumulative averages already for times $t=20$, which can be considered as a very loose upper bound for $\tau_{\rm prethermal}$.

\section{Computational details for perturbative currents}
For exact calculation, we use a second-ordered quantum circuit based Suzuki-Trotter algorithm with time step $dt=0.01$ whose details can be found in Ref.~\cite{XuPoletti2022}.    
To achieve large bath sizes, i.e., $\sim26$ per bath, we perform the perturbative calculations described in the main article. Taking Eq.~(4) in the main paper as an example, we rewrite it as 
\begin{align} 
  & \overrightarrow{\mathcal{C}}_{\rm L}(t,\tau)   \nonumber \\
  = &\bra{\psi_{\rm L} (\tau_{\rm prep})} U_{\rm L}^\dagger(t)\dBL U_{\rm L}(t-\tau) \BL U(\tau) \ket{\psi_{\rm L}(\tau_{\rm prep}})
 \label{eq: twotimecorrL3}
\end{align}
where $U_{\rm L}(t)$ is defined as the time evolution operator corresponding to the left Hamiltonian $\HL$, and $\ket{\psi_{\rm L}(\tau_{\rm prep})}$ is the initial state of the bath evolved for a time $\tau_{\rm prep}$.
Thus $\overrightarrow{\mathcal{C}}_{\rm L}(t,\tau)$ is computed by evolving the left bath, which we do exactly, also applying the quantum circuit based Suzuki-Trotter algorithm, up to $26$ spins. Similarly we evaluate Eqs.~(5,6). 
Once the correlations are computed, the current can be readily evaluated from Eq. (3) in the main paper. Note that it is also possible to perform numerical exact calculations via matrix product states to achieve simulations with large bath sizes. However, the fast growth of entanglement entropy, due to the thermalization dynamics, makes it extremely demanding.

\section{Magnitude of currents and energy differences between the baths \label{sec: energy_dependence}}

\begin{figure}
  \includegraphics[width=\columnwidth]{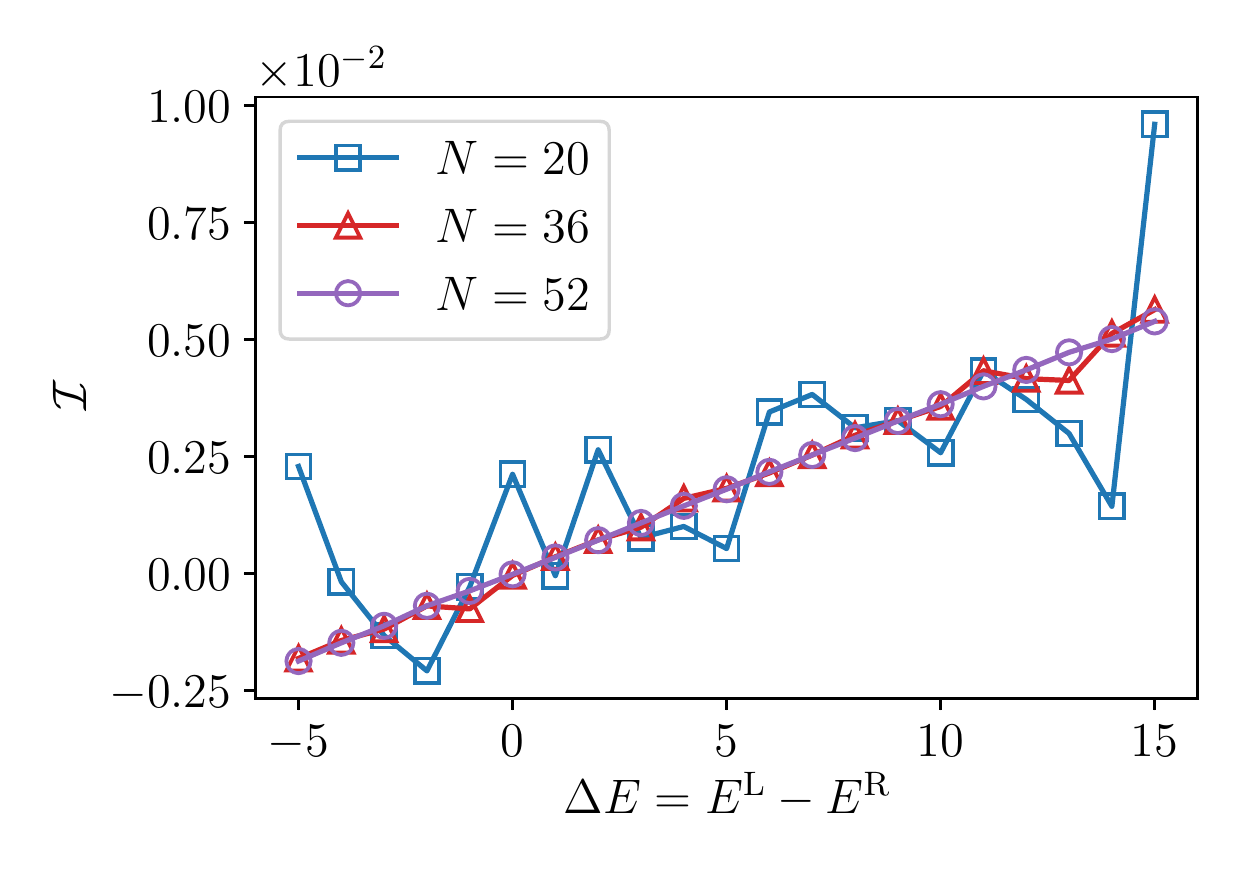}
  \caption{Current versus energy difference between left and right baths $\Delta E = \EL - \ER $ at $t=5$ for a single random realization of product initial state. The currents are computed via the perturbative expression given by Eq. (3) in the main paper.}

  \label{fig: energy}
\end{figure}

We have shown in the main paper that the typical current shows an exponentially decreasing variance with respect to the system size. This allows us to probe the properties of the system with simply a single `typical' realization provided the system size is large, i.e. $N \ge 20$. We show in Fig.~\ref{fig: energy} that for the parameters chosen, the quasi-steady current value at an already short time $t=5$ changes linearly with the energy difference between the two baths (vanishing when the energy difference approaches zero as expected), and the linear dependence becomes clearer as the size of the baths increases. For each data point, we consider one single random realization with the energy constraint, and each of these realizations starts with a different random seed. We remind the reader that for $N>28$ we use the perturbative approach which is, for the chosen magnitude of the coupling between the baths, extremely accurate.  

\section{Ultra-weak coupling between the baths}\label{sec: ultraweak}
When the coupling strength between the baths is even weaker than that in the main article, e.g., we now consider $\gamma=0.01$, one can observe that the variance of a single current realization is relatively larger when compared to the current value, so much that one can, temporarily, even witness current reversal. However, overall the time-averaged current matches the microcanonical current and, more importantly, as we have shown in Fig. 3 in the main article, when the bath sizes increase, we obtain an exponentially decreasing variance while the magnitude of current decreases linearly with $\gamma^2$, in short, still the emergence of strong prethermalization.

\begin{figure}
  \includegraphics[width=\columnwidth]{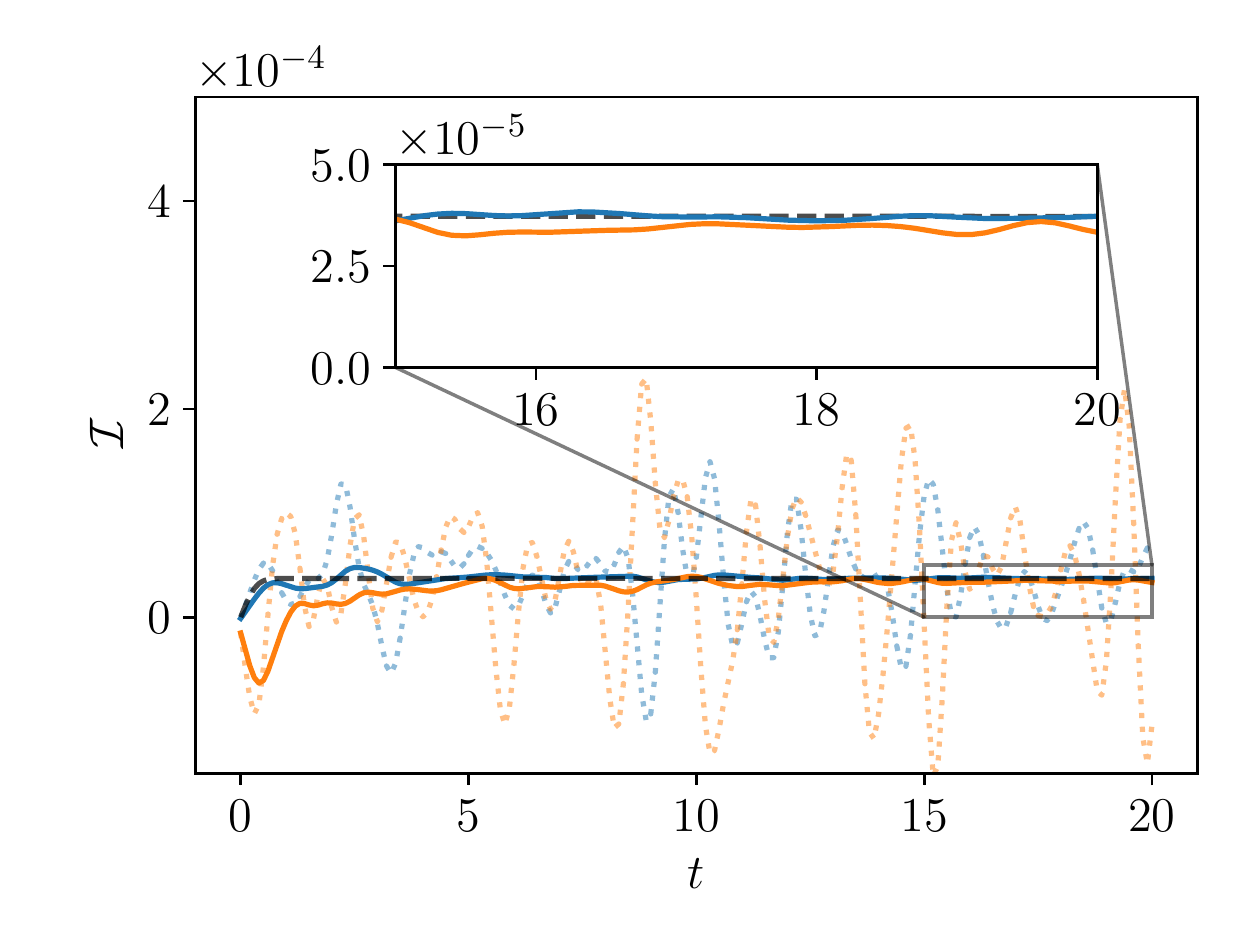}
  \caption{Current dynamics for exact (dotted lines) and time-averaged (solid lines) for $N=28$ (blue) and $N=24$ (orange) and $\gamma=0.01$. The dashed line denotes the perturbative microcanonical current for $N=24$.}\label{fig: ultraweak}
\end{figure}